\documentclass[aps,prd,twocolumn,tightenlines,showpacs,amsmath,amssymb,nofootinbib]{revtex4-1}

\usepackage{graphicx}
\usepackage{amsmath,amssymb}
\usepackage{bm}
\usepackage{epsfig}
\usepackage{color}              
\usepackage{comment}
\usepackage{hyperref}
\usepackage{array, float,tabularx}
\usepackage{lipsum}

\def\be{\begin{equation}}
\def\ee{\end{equation}}
\def\ba{\begin{eqnarray}}
\def\ea{\end{eqnarray}}
\def\bal{\begin{align}}
\def\eal{\end{align}}
\def\bald{\begin{aligned}}
\def\eald{\end{aligned}}

	\def \Tr {{\rm Tr}}
	\def \bs {\boldsymbol}
	\newcommand{\expo} [1] {{\rm e}^{ #1 }}

	\newcommand{\bigO} [1] {\mathcal{O} \left( #1 \right)}
	
	\renewcommand{\mod} [2] {#1~({\rm mod}~#2)}
	
\begin{document}

\title{On the representation (in)dependence of $k$-strings in pure Yang-Mills theory via supersymmetry}

\date{\today}

\author{Mohamed M. Anber}
\email{manber@lclark.edu}
\affiliation{Department of Physics, Lewis \& Clark College, 
Portland, OR 97219, USA}  
\author{Vito Pellizzani}
\email{vito.pellizzani@epfl.ch}
\affiliation{Institute de Th\' eorie des Ph\' enom\` enes Physiques, \' Ecole Polytechnique F\' ed\' erale de Lausanne, CH-1015 Lausanne, Switzerland}

\begin{abstract}
We exploit a conjectured continuity between super Yang-Mills on $\mathbb R^3\times \mathbb S^1$ and pure Yang-Mills to study $k$-strings in the latter theory. As expected, we find that Wilson-loop correlation functions depend on the N-ality of a representation ${\cal R}$ to the leading order. However, the next-to-leading order correction is not universal and is given by the group characters, in the representation ${\cal R}$, of the permutation group. We also study W-bosons in super Yang-Mills and show that they are deconfined on the string worldsheet, and therefore, can change neither the string N-ality nor its tension. This phenomenon mirrors the fact that soft gluons do not screen probe charges with non-zero N-ality in pure Yang-Mills.  Finally, we comment on the scaling law of $k$-strings in super Yang-Mills and compare our findings with strings in Seiberg-Witten theory, deformed Yang-Mills theory, and holographic studies that were performed in the 't Hooft large-$N$ limit. 
 \end{abstract}

\maketitle

\section{Introduction}

Flux tubes, or strings, are among the most fascinating objects in physics. They emerge as long-distance phenomena of various field theories, from abelian Higgs model to quantum chromodynamics (QCD). Although we have a good understanding of abelian strings (Abrikosov-Nielsen-Olesen type \cite{Abrikosov:1956sx,Nielsen:1973cs}), QCD strings remain poorly understood \cite{Greensite:2003bk,Shifman:2005eb}, thanks to the strong coupling of QCD. 

One of the important questions in Yang-Mills theories is how the string tension depends on the representation of the probe charges. The general lore, which is based on a pure physical argument, is that the string tension cannot depend on the representation. Instead, it can only depend on its N-ality. The N-ality of a representation ${\cal R}$ of $su(N)$ is defined as the number of  boxes in the Young tableau of  ${\cal R}$ modulo $N$. The physical argument in pure Yang-Mills goes as follows: since one can convert one representation ${\cal R}_1$ with N-ality $k$ to another representation ${\cal R}_2$ with the same N-ality by emitting soft gluons\footnote{The gluons are in the adjoint representation, and hence they have zero N-ality. Also, remember that in pure Yang-Mills there is no dynamical matter that can screen the probe charges.}, the string tension $\sigma_k$ will depend only on the N-ality $k$ and not on the representation. Unfortunately, it is extremely difficult to provide a direct mathematical proof of such an intuitive argument; the strong coupling nature of QCD hinders the chances to find such proof.

Lattice field theory provides a nonperturbative definition of strongly coupled theories, and therefore, one hopes that direct simulations of Yang-Mills theory can provide complete nonperturbative pictures of QCD strings. Practical lattice simulations of QCD, however, suffer from lattice artifacts leading to some dependence of the string tension on the representation \cite{Gliozzi:2005dv,Campbell:1985kp,Poulis:1995nn}, which is particularly evident in the case of a large number of colors. This is because the relaxation time of higher representation strings can be exponentially large, which mistakenly can signal a dependence of the string tension on the representation rather than its N-ality. Lattice strong coupling expansion, in addition, suffers from the same artifact \cite{Armoni:2006ri}.

Fortunately, the AdS/CFT correspondence can shed some light on the question at hand. In particular, it was shown in \cite{Armoni:2006ri} (also see \cite{Gross:1998gk}) that the expectation value of the Polyakov's Loop in a representation ${\cal R}$ is given by $\langle {\cal P}_{\cal R}\rangle=F({\cal R})e^{-\sigma_k A}$, where $A$ is the area of the Polyakov's loop. Thus, as expected, the string tension depends only on the N-ality $k$, while there is a nonuniversal representation dependent prefactor $F({\cal R})$. This behavior, however, was shown only in the 't Hooft large-$N$ limit, leaving behind the finite $N$ case with no direct answer. 

\begin{figure}[t]
\begin{center}
\includegraphics[width=.5\textwidth]{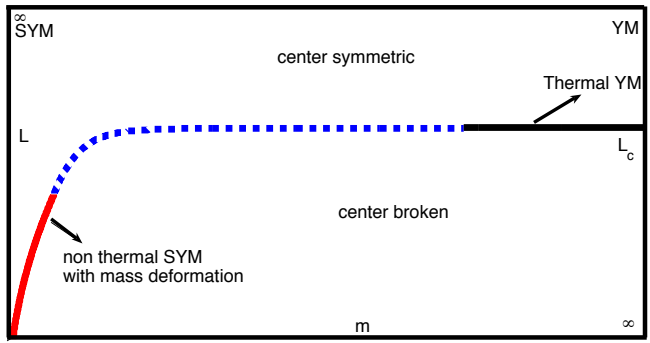}
\end{center}
\caption{Continuity between mass deformed ${\cal N}=1$ super Yang-Mills on $\mathbb R^3 \times \mathbb S^1$ and pure Yang-Mills at finite temperature. The red thick curve in the lower left corner is the phase separation between the center symmetric and center broken phases in super Yang-Mills on $\mathbb R^3\times \mathbb S^1$. This part of the phase diagram is under analytical control since the theory is in its weakly-coupled semi-classical regime. The black curve in the upper right corner is the phase separation between the confined and deconfined phases of the strongly-coupled pure Yang-Mills. This part of the curve can be envisaged using lattice Monte Carlo simulations. The dotted curve is conjectured to be smoothly connecting both the weakly coupled and strongly coupled theories.}
\label{conjecture}
\end{figure}

The lack of a direct proof of the expected universality of string tension, specially for finite $N$, calls for a new perspective on the problem. A novel way to approach strongly coupled pure Yang-Mills is to exploit a conjectured continuity that first appeared in \cite{Poppitz:2012sw}. This is a continuity between softly broken (via a mass term) ${\cal N}=1$ super Yang-Mills on $\mathbb R^3 \times \mathbb S^1$, where $\mathbb S^1$ is a spatial rather than a thermal circle, and pure Yang-Mills at finite temperature. According to this continuity, the quantum phase transition in the former theory is continuously connected to the thermal phase transition in the latter one.  This is illustrated in FIG. \ref{conjecture}. At small circle circumference $L$ and small gaugino mass $m$ (this is the lower left corner, the red curve, of FIG. \ref{conjecture}) the theory is confining, in a weakly coupled regime, has a preserved $\mathbb Z_N$ center symmetry,  and is under complete analytical control. Therefore, by varying $m$ or $L$ the theory experiences a quantum phase transition and one goes from a center-symmetric phase (at small $m$ and $L$) to a center-broken phase (larger values of $m$ and $L$).  On the other hand, as $m \rightarrow \infty$ the gaugino decouples and the theory flows to a pure Yang-Mills over $\mathbb S^1$ (the right side in FIG. \ref{conjecture}). This is a pure Yang-Mills theory \footnote{In the limit $m \rightarrow \infty$ there is no dynamical matter. Hence, the fact that we started with a spatial , rather than a thermal, circle does not make any difference, since the gauge fluctuations always obey periodic boundary conditions.} at finite temperature $T=1/L$. This is a strongly coupled theory whose phase transition can only be inferred from strong coupling calculations, e.g., lattice simulations. According to the continuity conjecture in  \cite{Poppitz:2012sw}, the quantum phase transition in super Yang-Mills is continuously connected to the thermal phase transition in pure Yang-Mills. This continuity is indicated by the dashed line in the intermediate region in FIG. \ref{conjecture}. Despite the fact that a proof of the continuity is still lacking, many checks have shown that various physical observables share the same qualitative behavior in both limits $m\rightarrow 0$ and $m \rightarrow \infty$. This includes the nature of phase transition, i.e. first versus second order \cite{Poppitz:2012sw,Poppitz:2012nz,Anber:2014lba,Poppitz:2013zqa}, the dependence of the critical temperature on the $\theta$ angle \cite{Anber:2013sga}, and the dependence of the fundamental string on temperature \cite{Anber:2014lba}.

In the present note we push the continuity even further: we check whether {\em correlation functions in the mass deformed ${\cal N}=1$ super Yang-Mills on $\mathbb R^3\times S^1$ and in pure Yang-Mills are continuously connected}. This demands that {\em correlation functions do not experience a phase transition as long as we do not cross the phase separation line in FIG. \ref{conjecture}}.  The validity of this conjecture as well as its limitations is the main subject of the present work. If this continuity holds, then it can provide a new venue to analytically study various observables, including the strings, which are otherwise very hard to compute directly in the strongly coupled theory.

\begin{figure}[t]
\begin{center}
\includegraphics[width=84mm]{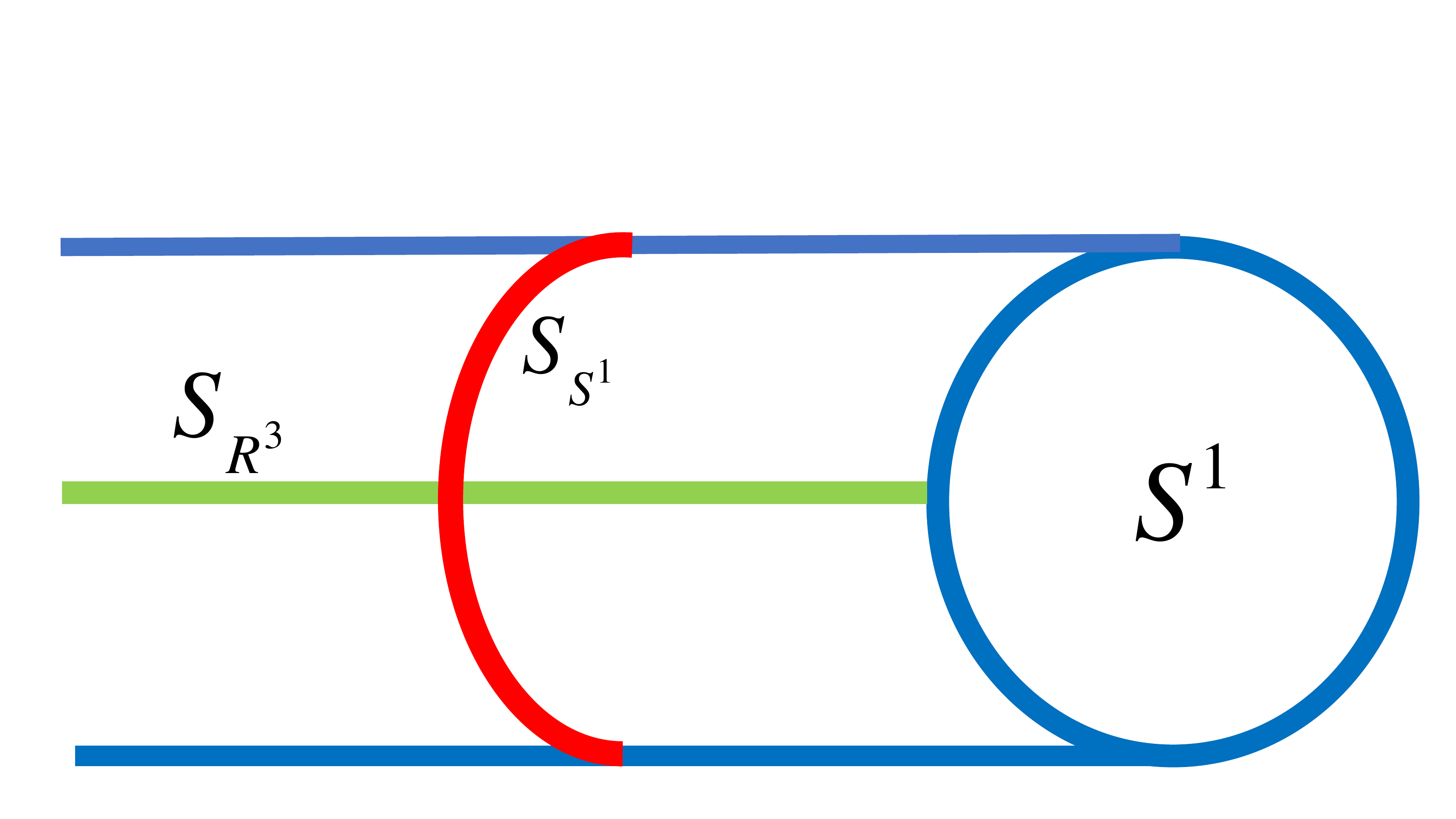}
\caption{There are two types of strings in super Yang-Mills on $\mathbb R^3 \times \mathbb S^1$. The first type (the green color line), which we denote by ${\cal S}_{\mathbb R^3}$, is the string between two prob charges located on the $\mathbb R^2$ plane. The other type of strings (the red color), which we denote by ${\cal S}_{\mathbb S^1}$, wraps around the $\mathbb S^1$ circle. It is this second type of strings that can be interpreted as pure Yang-Mills k-strings in the limit $m \rightarrow \infty$. }
\label{strings types}
\end{center}
\end{figure}

There are two types of strings in super Yang-Mills on $\mathbb R^3\times  \mathbb S^1$: the strings on $\mathbb R^3$ between two probe charges located on the $\mathbb R^2$ plane, which we denote\footnote{The ${\cal S}_{\mathbb R^3}$ strings in deformed Yang-Mills theory are thoroughly studied in \cite{Poppitz:2017ivi}. A similar study of ${\cal S}_{\mathbb R^3}$ in super Yang-Mills is left for a future work.} by ${\cal S}_{\mathbb R^3}$, and the strings that wrap around the circle $\mathbb S^1$, which we denote by ${\cal S}_{\mathbb S^1}$. According to the continuity picture, the ${\cal S}_{\mathbb S^1}$ strings are the ``would be'' k-strings in pure Yang-Mills theory in the limit $m\rightarrow \infty$; the $\mathbb S^1$ circle (which is a space-like circle) becomes the thermal circle in pure Yang-Mills in the decoupling limit. This picture is depicted in FIG.\ref{strings types}.

 In particular, in this work we calculate the tension of these ``would be'' k-strings in pure Yang-Mills theory.  This is carried out by computing the Polyakov-Loop correlator in super Yang-Mills  deep in the weak-coupling confining regime. This is the Polyakov-loop that wraps around the spatial $\mathbb S^1$ circle: ${\cal P}_{\cal R}=\mbox{Tr}_{\cal R}\exp\left[i{\oint_{\mathbb S^1}\bm A_{3}}\right]$, where $\bm A_3$ is the gauge field component along the circle and the trace is take in representation ${\cal R}$. Because the theory is in a gapped phase, then for a very large separation between two Polyakov's loops one has $\mbox{lim}_{r\rightarrow \infty}\langle {\cal P}_{\cal R}(\bm 0){\cal P}_{\cal R}^\dagger(\bm r)\rangle={\cal F}_{\cal R} e^{-\sigma_{\cal R} rL}$, where $\sigma_{\cal R}$ is a constant that can be exactly determined since the theory is in a calculable regime.  According to the conjectured continuity $\sigma_{\cal R}$ should correspond to the string tension in pure Yang-Mills that also wraps around $\mathbb S^1$. Thus, by computing the trace in any representation ${\cal R}$, one can infer the dependence of the string tension on ${\cal R}$. Our calculations show that for any finite $N$ the string tension $\sigma_{{\cal R}}$ depends, to leading order, on the N-ality of the representation ${\cal R}$. The pre-coefficient ${\cal F}_{\cal R}$, however, is found to depend on the representation.  Albeit in a weakly coupled regime, this is the first direct proof of the leading-order independence of the string tension of its representation for finite $N$.  

Our work is organized as follows. In Section \ref{Mass deformed super Yang-Mills} we review the basics of mass deformed super Yang-Mills on $\mathbb R^3 \times \mathbb S^1$ and set up the notation and convention. Since this topic has been studied in great details in the literature, we only provide the necessary formalism that enable the reader to grasp the main ideas. The main results of this section are Eqs. \ref{mass eigenvalues}, \ref{p space correlator}, and \ref{real space correlator}. Experts can skip this section to Section \ref{Polyakov's loop correlator and string tension}, where we provide a direct proof that the Polyakov-loop correlator depends, to leading order, on the N-ality of a representation. In Section \ref{W-bosons on the string worldsheet} we study the W-bosons on the string worldsheet of super Yang-Mills and show that these bosons are deconfined on the string, and therefore, they cannot affect the string tension or its N-ality. Finally in Section  \ref{Discussion},  we  comment on the scaling of the ${\cal S}_{\mathbb S^1}$ strings and their large-N limit and we compare our findings with strings in Seiberg-Witten and deformed Yang-Mills theories. 
%

\section{Mass deformed super Yang-Mills}
\label{Mass deformed super Yang-Mills}

We consider ${\cal N}=1$ super Yang-Mills theory on $\mathbb R^3 \times \mathbb S^1$. This is an $su(N)$ Yang-Mills theory endowed with a single adjoint Weyl fermion (gaugino) obeying periodic boundary conditions along the circle $\mathbb S^1$. If we take the circumference of the circle, $L$, to be much smaller than the strong scale of the theory $\Lambda$, i.e. $N\Lambda L\ll 1$, then the theory enters its weakly coupled regime and becomes amenable to semi-classical treatment. Upon dimensionally reducing from $3+1$ to $3$ dimensions the theory generates a scalar field, which is the Wilson line holonomy along the circle: $\bm \Phi=\int_{\mathbb S^1}\bm A_3$. Supersymmetry guarantees the vanishing of the perturbative potential $V(\bm \Phi)$ that results from integrating out the tower of massive Kaluza-Klein excitations of gauge bosons and gauginos. Thus, the theory has a perturbatively exact flat direction such that turning on any non-zero value of $\bm \Phi$  causes the breaking of $su(N)$ to the maximum abelian torus $u(1)^{N-1}$.  In three dimensions the photons are dual to scalars, and hence, the $3$-D long-distance effective field theory contains massless scalars and fermions not charged under $u(1)^{N-1}$. The action of the theory reads:
\begin{eqnarray}
\nonumber
S&=&\frac{1}{L}\int d^3 x\left\{-\frac{1}{g^2}\left(\partial_\mu\bm \Phi\right)^2-\frac{g^2}{16\pi^2}\left(\partial_\mu\bm\sigma\right)^2  \right.\\
&&\left.-i\frac{2L^2}{g^2}\bar {\bm\lambda}\bar \sigma^\mu\partial_\mu\bm \lambda\right\}\,,
\end{eqnarray}
where $g$ is the four-dimensional coupling which is kept small, $\bm \sigma$ are the dual photons, and $\bm \lambda$ are the fermions. All light fields have components only along the Cartan generators $\bm H=\left(H_1,H_2,...,H_{N-1}\right)$, which are denoted by bold face letters, e.g., $\bm \sigma=\left(\sigma_1,\sigma_2,...,\sigma_{N-1}\right)$.

The story does not end at the perutrabative sector. The theory, in addition, admits nonperturbative saddles. These are the monopole-instantons which lift the flat direction and generate masses for the photons. The details of the story can be found in \cite{Davies:1999uw,Davies:2000nw,Unsal:2007jx,Poppitz:2012sw,Anber:2014lba}. In essence, the monopole-instantons generate the superpotential:
\begin{eqnarray}
{\cal W}\sim \sum_{a=1}^{N}e^{\bm \alpha_a\cdot \bm X+2\pi i\tau\delta_{a,N}}\,,
\end{eqnarray}
where $\bm X$ is the chiral multiplet,  $\tau=i\frac{4\pi^2}{g^2}+\frac{\theta}{2\pi}$, and $\theta$ is the vacuum angle. The sum is over the simple roots $\{\bm \alpha_a\}$, $a=1,2,..,N-1$ as well as the affine root $\bm\alpha_N=-\sum_{a=1}^{N-1}\bm \alpha_a$. The inclusion of the affine root is a crucial ingredient in order for the theory to have a stable vacuum. In fact, including this root in the sum is how the theory remembers its four dimensional origin and, as we will see, is responsible in a direct way for the observation that the string tension depends only on the N-ality of the representation to the leading order.  

The superpotential will generate the scalar potential (we call it the bion potential\footnote{Magnetic and neutral bions are correlated events made of two monopoles, which appear as a direct sequence of ${\cal K}^{i\bar j}\frac{\partial {\cal W}}{\partial X^i}\frac{\partial \bar{\cal W}}{\partial X^{\dagger j}}$,  see \cite{Anber:2011de,Poppitz:2012sw} for details.})  $V_{\mbox{\scriptsize bion}}={\cal K}^{i\bar j}\frac{\partial {\cal W}}{\partial X^i}\frac{\partial \bar{\cal W}}{\partial X^{\dagger j}}$, where ${\cal K}^{i\bar j}$ is the K{\"a}hler potential, which to zeroth order in the coupling constant $g$ is given\footnote{The one-loop correction to the K{\"a}hler potential was worked out in \cite{Anber:2014lba}. This correction becomes important only if the gauge group does not have a center, e.g., $G_2$. See \cite{Anber:2014lba} for details.} by ${\cal K}^{i\bar j}=\delta^{ij}$. 
  As we mentioned in the introduction, we also turn on a small gaugino mass which breaks the supersymmetry softly and generates a perturbative potential\footnote{The perturbative potential is the one-loop contribution from the Kaluza-Klein tower of gauge bosons and massive gauginos.}. In addition,  the gaugino mass lifts the monopole-instanton zero modes and give an additional contribution to the scalar potential $V_m$. 
	
 The supersymmetric theory, in the small $m$ and $L$ regime, has a preserved center symmetry and  the vacuum expectation value of the Wilson line holonomy is $\bm\Phi_0=\frac{2\pi}{N}\bm \rho$, where $\bm \rho=\sum_{a=1}^{N-1}\bm \omega_a$ is the Weyl vector, and $\bm \omega_a$ are the fundamental weights. Now we write 
\begin{eqnarray}
\bm \Phi=\bm \Phi_0+\frac{g^2}{4\pi^2}\bm b\,,
\end{eqnarray}
such that $\bm b$ are the small fluctuations of the adjoint scalar about the vacuum. 
 After taking the monopoles and gaugino mass into account, we find that  the total bosonic Lagrangian in terms of $\bm \sigma$ and $\bm b$ is given by 
\begin{eqnarray}
\nonumber
{\cal L}&=&\frac{1}{12\pi}\frac{m_W}{\log(m_W/\Lambda)}\left(\left(\partial_\mu \bm b\right)^2+\left(\partial_\mu \bm \sigma\right)^2\right)\\
&&+V_{\mbox{\scriptsize np}}+V_{\mbox{\scriptsize pert}}\,,
\label{main lagrangian}
\end{eqnarray}
where $V_{\mbox{\scriptsize np}}$ and $V_{\mbox{\scriptsize pert}}$ are respectively the nonperturbative and perturbative potentials and $m_W=\frac{2\pi}{NL}$ is the W-boson mass. As was shown in \cite{Poppitz:2012sw},  $V_{\mbox{\scriptsize pert}}$ is suppressed by three powers of $\log(m_W/\Lambda)$ compared to  $V_{\mbox{\scriptsize np}}$, and hence, we neglect it in our analysis. The nonperturbative potential contains contributions from two parts: (1) the monopole part, which is non-vanishing if and only if the gauginos are massive (massless gauginos have two zero modes in the background of monopoles, and hence, the latter cannot contribute to the bosonic potential), and (2) the magnetic and neutral bion potential, see footnote 4.  The nonperturbative potential is given by
\begin{eqnarray}
\nonumber
V_{\mbox{\scriptsize np}}&=&V^0_{\mbox{\scriptsize bion}}\left[\sum_{a=1}^{N}e^{-2\bm \alpha_a\cdot \bm b}-e^{-\left(\bm \alpha_a+\bm\alpha_{a+1}\right)\cdot \bm b}\right.\\
\nonumber
&&\left.\quad\quad\quad\quad\quad\quad\quad\quad\quad\times\cos\left[\left(\bm\alpha_a-\bm\alpha_{a+1}\right)\cdot \bm\sigma \right] \right]\\
&&-V_{\mbox{\scriptsize mon}}^0\left[\sum_{a=1}^{N}e^{-\bm \alpha_a\cdot \bm b}\cos\left[\bm\alpha_a\cdot \bm\sigma+\psi \right]\right]\,,
\label{np potential}
\end{eqnarray}
where $\psi=\frac{2\pi \ell+\theta}{N}$, and the parameter $\ell=0,1,..,N-1$ labels the vacuum branch, i.e. the branch with minimum ground energy. The  bion and monopole coefficients $V^0_{\mbox{\scriptsize bion}}$ and  $V^0_{\mbox{\scriptsize mon}}$, expressed in terms of the physical mass $m_W$ and the strong scale  $\Lambda$, are given by
\begin{eqnarray}
\nonumber
V^0_{\mbox{\scriptsize bion}}&=&\frac{27}{8\pi} \frac{\Lambda^6}{m_W^3}\log \left(\frac{m_W}{\Lambda}\right)\,,\\
V^0_{\mbox{\scriptsize mon}}&=&\frac{9}{2\pi}\frac{m\Lambda^3}{m_W}\log\left(\frac{m_W}{\Lambda}\right)\,.
\end{eqnarray} 
For convenience, we also introduce the dimensionless gaugino mass parameter 
\begin{eqnarray}
c_m=\frac{V^0_{\mbox{\scriptsize mon}}}{V^0_{\mbox{\scriptsize bion}}}=\frac{4mm_W^2}{3\Lambda^3}=\frac{16\pi^2m}{3\Lambda (\Lambda LN)^2}\,.
\end{eqnarray}

To further proceed, one needs to find the masses of the fluctuations $\bm b$. Expanding $V_{\mbox{\scriptsize np}}$ to quadratic order in $\bm b$ and $\bm \sigma$ and rescaling $\bm b$ and $\bm \sigma$ as $\{ b_a^2,\sigma_a^2\} \rightarrow \frac{6\pi\log\left(m_W/\Lambda\right)}{m_W} \{b_a^2,\sigma_a^2\} $ to have a canonically normalized Lagrangian, we obtain:
\begin{eqnarray}
{\cal L}&=&\frac{1}{2}\left(\partial_\mu \bm b\right)^2+\frac{1}{2}\left(\partial_\mu \bm \sigma\right)^2+{\cal V}\,,
\end{eqnarray}
and 
\begin{eqnarray}
\nonumber
{\cal V}&=&-N c_m\cos\psi\\
\nonumber
&&+  m_0^2\sum_{a=1}^{N}\left[\left(\bm \alpha_a\cdot \bm b\right)^2-\left(\bm \alpha_{a+1}\cdot \bm b\right)\left(\bm \alpha_{a}\cdot \bm b\right) \right.\\
\nonumber
&&\left.\quad\quad\quad+ \left(\bm\alpha_a\cdot \bm \sigma\right)^2-\left(\bm \alpha_{a+1}\cdot \bm \sigma\right)\left(\bm \alpha_{a}\cdot \bm \sigma\right)  \right.\\
\nonumber
&&\left.\quad\quad\quad+\frac{c_m}{2}\left( \left(\bm\alpha_a\cdot \bm \sigma\right)^2-\left(\bm\alpha_a\cdot \bm b\right)^2  \right)\cos\psi\right.\\
&&\left.\quad\quad\quad-c_m\left(\bm\alpha_a\cdot \bm \sigma\right)\left(\bm\alpha_a\cdot \bm b\right)\sin\psi\right]\,,
\label{V before diagonalizing}
\end{eqnarray} 
where 
\begin{eqnarray}
m_0^2=\frac{81}{4}\frac{\Lambda^6\left[\log(m_W/\Lambda) \right]^2}{m_w^4}\,.
\label{photon masses}
\end{eqnarray}

 The easiest way to obtain the mass spectra of the quadratic Lagrangian is to go to the $\mathbb R^N$ root basis. In this basis the weights of the fundamental representations are given by
\begin{eqnarray}
\bm \nu_a=\bm e_a-\frac{1}{N}\,,a=1,2,...,N\,,
\end{eqnarray}
while the roots are 
\begin{eqnarray}
\nonumber
&&\left\{\bm \alpha_a=\bm e_a-\bm e_{a+1}, 1\leq a\leq N-1\,,\right.\\
&&\left.\bm  \alpha_N=\bm e_{N}-\bm e_1\right\}\,.
\end{eqnarray}
Notice the cyclic structure of the roots in these basis. Also, notice that the affine root $\bm \alpha_N$ is the link that completes the cycle.

The cyclic nature of $\{\alpha_a\}$, $a=1,..,N$ enable us to use the discrete Fourier transform defined by:
\begin{eqnarray}
\left\{\begin{array}{c} b_j\\ \sigma_j \end{array}  \right\}=\frac{1}{\sqrt{N}}\sum_{p=0}^{N-1} \left\{\begin{array}{c} \tilde b_p\\ \tilde\sigma_p \end{array}  \right\}e^{-2\pi i \frac{p_j}{N}}\,.
\end{eqnarray}
In doing so, we have introduced the fictitious degree of freedom $b_0$, the zero mode, which decouples from the rest of excitations. 
Had we not included the monopole corresponding to the affine root, we would not be able to use the discrete Fourier transform to simplify our calculations. As we will see in the next section, this transform is pivotal in our proof of the N-ality dependence of the string tension. Now, we substitute the $\mathbb R^N$ root vectors into Eq. \ref{V before diagonalizing} and use the discrete Fourier transform to find after straightforward algebra
\begin{eqnarray}
\nonumber
{\cal V}&=&-N c_m \cos \psi\\
&&+m_0^2\sum_p {\cal A_-} \tilde b_p \tilde b_{-p}+{\cal A_+} \tilde\sigma_p \tilde\sigma_{-p}+{\cal C}\tilde\sigma_p \tilde b_{-p}\,,
\end{eqnarray}
where
\begin{eqnarray}
\nonumber
{\cal A}_\pm&=&8\sin^4\left(\frac{\pi p}{N}\right)\pm 2c_m\sin^2\left(\frac{\pi p}{N}\right)\cos\psi\,,\\
{\cal C}&=&-4c_m \sin^2\left(\frac{\pi p}{N}\right)\sin\psi\,.
\end{eqnarray} 

In order to further decouple $\tilde\sigma_p$ and $\tilde b_p$, we define new fields $\tilde \sigma'_p$ and $\tilde b'_p$:
\begin{eqnarray}
\nonumber
\tilde b'_p&=&\cos\frac{\psi}{2}\tilde b_p+\sin \frac{\psi}{2}\tilde \sigma_p\,,\\
\tilde \sigma'_p&=&-\sin\frac{\psi}{2}\tilde b_p+\cos\frac{\psi}{2}\tilde \sigma_p\,.
\end{eqnarray}
The mass square eigenvalues of $\tilde \sigma'_p$ and $\tilde b'_p$ are
\begin{eqnarray}
\nonumber
{\cal M}_{\tilde \sigma'_p}^2&=&16m_0^2\left[\sin^4\left(\frac{p\pi}{N}\right)+ \frac{c_m}{4} \sin^2\left(\frac{p\pi}{N}\right) \right]\,,\\
{\cal M}_{\tilde b'_p}^2&=&16m_0^2\left[\sin^4\left(\frac{p\pi}{N}\right)- \frac{c_m}{4} \sin^2\left(\frac{p\pi}{N}\right) \right]\,,
\label{mass eigenvalues}
\end{eqnarray}
where $p=1,2,...,N-1$, and we neglected the zero mode $p=0$. 

Now we are in a position to calculate the correlator  $\langle \tilde b_p(\bm 0) \tilde b_{-p}(\bm r) \rangle$. We consider the Euclidean version of our theory such that  $\bm r$ is a three dimensional vector (the Euclidean time is taken along the third direction). Keeping in mind that the fields $\tilde \sigma'_p$ and $\tilde b'_p$ do not couple, we find that the propagator $\langle \tilde b_p(\bm 0) \tilde b_{-p}(\bm r) \rangle$ is given by:
\begin{widetext}
\begin{eqnarray}
\langle \tilde b_p(\bm 0) \tilde b_{-p}(\bm r) \rangle=\cos^2\frac{\psi}{2}\langle \tilde b'_p(\bm 0)\tilde b'_{-p}(\bm r) \rangle+\sin^2\frac{\psi}{2}\langle \tilde \sigma'_p(\bm 0)\tilde \sigma'_{-p}(\bm r) \rangle
=\frac{1}{4\pi r}\left\{\cos^2\frac{\psi}{2}e^{-{\cal M}_{\tilde b'_p}r}+\sin^2\frac{\psi}{2}e^{-{\cal M}_{\tilde \sigma'_p}r}  \right\}\,.
\label{p space correlator}
\end{eqnarray}
\end{widetext}
In sequence, we use the inverse discrete Fourier transform to obtain the correlator 
\begin{eqnarray}
\langle  b_j(\bm 0)  b_{l}(\bm r) \rangle=\frac{1}{N}\sum_{p=0}^{N-1}e^{-\frac{2\pi i p}{N}(j-l)} \langle \tilde b_p(\bm 0) \tilde b_{-p}(\bm r) \rangle\,.
\label{real space correlator}
\end{eqnarray}
The exponents of  the correlator $\langle  b_j(\bm 0) b_{k}(\bm r) \rangle$ are independent of  \footnote{Thus, we need to go to the next to leading order correction in $g$ to find the dependence of the string tension on $\theta$; see \cite{Unsal:2012zj, Thomas:2011ee, Anber:2014lba}.} $\theta$. From here on, we set $\theta=0$ and select the vacuum branch $\ell=0$. Therefore, the correlator $\langle  b_j(\bm 0) b_{k}(\bm r) \rangle$ receives a contribution only from the first term in (\ref{p space correlator}). We note that the masses ${\cal M}_{\tilde b'_p}$ are much lighter than the W-boson mass, $\frac{\pi}{NL}$, as can be checked from (\ref{photon masses}). The string ${\cal S}_{\mathbb S^1}$ that wraps around $\mathbb S^1$ is made of the light excitations ${\cal M}_{\tilde b'_p}$, and therefore, the string thickness $\sim {\cal M}_{\tilde b'_p}^{-1}$ is much bigger than the circle $\mathbb S^1$. This fact is responsible for the square sine scaling of the ${\cal S}_{\mathbb S^1}$ string, as we discuss in the conclusion.

\section{Polyakov's loop correlator and string tension}
\label{Polyakov's loop correlator and string tension}

In this section we use the conjectured continuity between mass deformed ${\cal N}=1$ super Yang-Mills and pure Yang-Mills to show that the string tension of the latter theory depends only on the N-ality of the representation to the leading order. In order to show that we visualize the Polyakov's loop along the $\mathbb S^1$ circle $\Tr_{\cal R}\exp\left[i\oint_{\mathbb S^1}\bm A_3\right]$  as a string wrapping the circle. We can calculate the correlator of two Polyakov's loop in the small $L$ and $m$ regime, where the theory is confining, has a preserved center symmetry, weakly coupled, and under analytical control. We prove that the correlator  $\mbox{lim}_{r\rightarrow \infty}\langle {\cal P}_{\cal R}(\bm 0){\cal P}_{\cal R}^\dagger(\bm r)\rangle={\cal F}_{{\cal R}} e^{-\sigma_{\cal R} r}$, where $\sigma_{\cal R}$ is a constant that depends only on the N-ality of the representation ${\cal R}$ and the pre-factor ${\cal F}_{{\cal R}}$ depends on the representation ${\cal R}$. Then, by continuity (the absence of phase transitions as we take the gaugino mass to infinity), we argue that $\sigma_{\cal R}$ can be interpreted as the string tension of a pure Yang-Mills theory that depends only on its N-ality, as expected on physical grounds. 

\subsection{From the fundamental to any representation of $su(N)$}

We first summarize a few important results from group theory concerning traces of $su(N)$ elements in general representations. The following discussion holds for any $N \geq 3$. Let ${\cal R}=(y_1, y_2, \dots, y_{N-1})$ denote the Young tableau with $y_i$ columns of $i$ boxes (where bigger columns are placed on the left as usual), which is associated with a particular tensor representation ${\cal R}$ of $su(N)$.  
Now, let $P$ be an element in $su(N)$. The trace of $P$ in a general representation ${\cal R}$ can be written as a sum of products of fundamental traces as is given by the Frobenius formula (see \cite{Anber:2017pak} and references therein): 
\begin{eqnarray}
\nonumber
\mbox{Tr}_{\cal R}P&=&\frac{1}{n!}\sum_{\vec j\in S_n}\chi_{\cal R}\left(\vec j\right)\left(\mbox{Tr}_F P\right)^{j_1}\left(\mbox{Tr}_F P^2\right)^{j_2}\\
&&\times...\left(\mbox{Tr}_F P^n\right)^{j_n}\,,
\label{Frobenius formula}
\end{eqnarray}
where $n$ is the number of boxes in Young tableau of representation ${\cal R}$  (not mod $N$) and $S_n$ is the permutation group. The permutations $\vec{j} = \{j_1, \cdots, j_n\} \in S_n$ are most easily found as solutions of $\sum_{k = 1}^n k  j_k = n$. For example, for $S_2$ we have $\vec j = \left\{ (2, 0), (0, 1) \right\}$ and for $S_3$ we have $\vec j = \left\{ (3, 0, 0), (1, 1, 0), (0, 0, 1) \right\}$, etc. $\chi_{\cal R}\left(\vec j\right)$ is the group character, in the representation {\cal R}, of the permutation $\vec j$. This sets the ground to obtaining the Polyakov-loop correlator in any representation ${\cal R}$ in terms of the fundamental representation. We will show that the correlator, to leading order, depends only on the N-ality of the representation and not on the representation itself.

\subsection{Perturbative expansion of the Polyakov's loop correlator}

We now turn to the derivation of the Polyakov's loop correlator in a general representation ${\cal R}$ of $su(N)$. Since our effective field theory is valid to zeroth-loop order,  we shall focus on the correlator expansion up to $\bigO{g^4}$ in the coupling constant.  The Wilson line operator reads:
\begin{equation}
	\Omega(\bm r) =\exp\left[\oint_{\mathbb S^1}i\bm A_3\right] =\expo{i \bm{H} \cdot \bm{\Phi}(\bm r)}\,,
\end{equation}
with $\bm r$ being a three dimensional Euclidean vector and the Wilson line wraps the $\mathbb S^1$ circle. For $su(N)$, the vacuum is given by $\bs{\Phi}_0 = \frac{2 \pi}{N} \bs{\rho}$. As we pointed out in Section \ref{Mass deformed super Yang-Mills}, we write 
\begin{eqnarray}
	\bm{\Phi} = \bs{\Phi}_0 + \frac{g^2}{4 \pi} {\bm{b}}\,.
\end{eqnarray}
We are interested in the Polyakov-loop correlator in representation ${\cal R}$:
\begin{equation}
\langle {\cal P}_{\cal R}(\bm 0) {\cal P}_{\cal R}^\dagger(\bm r)\rangle\equiv\langle \Tr_{\cal R} \Omega(0)  \Tr_{\cal R} \Omega^\dagger(\bm r) \rangle.
\end{equation}
This is the correlator between two Polyakov's loops wrapping around the $\mathbb S^1$ circle and located at $0$ and $\bm r$.
To this end, let us pick any $k \neq \mod{0}{N}$ and define $\Omega_0^k \equiv \expo{i k \bs{H} \cdot \bs{\phi}_0^{(0)}}$, whose eigenvalues are evenly spread around the unit circle, whence, $\Tr_F \Omega_0^k = 0$. We now expand in powers of $\bm b$ the trace in the fundamental of the $k$-th power of Polyakov's operator (recall that the Cartan generators $\bm H$ commute and that $\bm b$ are small fluctuations of the holonomy field about the vacuum):
\begin{widetext}
\begin{eqnarray}
\nonumber
	\Tr_F \Omega^k(\bm r)  &=&   \Tr_F \left[ \Omega_0^k \exp \left( \frac{i k g^2}{4 \pi} \bs{H} \cdot {\bs{b}}(\bm r)\right) \right]
	\cong \Tr_F \left[ \Omega_0^k \left\{ \bm 1 + \frac{i k g^2}{4 \pi} \bs{H} \cdot \tilde{\bs{b}}(\bm r)
+ \frac{g^4}{32 \pi^2}  \left( i k \bs{H} \cdot {\bs{b}}(\bm r) \right)^2 \right\} \right] + \bigO{g^6} \\
	 &=& \frac{i k g^2}{4 \pi} B_k(\bm r) + \frac{g^4}{32 \pi^2} C_k(\bm r) + \bigO{g^6}\,,
\end{eqnarray}
\end{widetext}
where 
\begin{eqnarray}
\nonumber
B_k(\bm r) &\equiv& \Tr_F \left[ \Omega_0^k \bs{H} \cdot {\bs{b}}(\bm r) \right]\,,\\
C_k(\bm r) &\equiv& \Tr_F \left[ \Omega_0^k  \left( i k \bs{H} \cdot {\bs{b}}(\bm r) \right)^2\right]\,.
\end{eqnarray}
Moreover, since there is no $\bigO{g^0}$ term, we further obtain
\begin{eqnarray}
	\left( \Tr_F \Omega^k(\bm r) \right)^2 = - \frac{k^2 g^4}{16 \pi^2} B_k^2(\bm r) + \bigO{g^6}\,,
\end{eqnarray}
and $\left( \Tr_F \Omega^k(\bm r) \right)^a \sim \bigO{g^6}$ for $a > 2$.

 Now, let us make use of the Frobenius formula (\ref{Frobenius formula}). For a representation ${\cal R}$ of $su(N)$, corresponding to a Young tableaux of $n$ boxes (not mod $N$), express $\Tr_{\cal R} \Omega(x)$ in terms of $\Tr_F$ and expand in $g$. The only $\bigO{g^2}$ contribution in this expansion comes from the term with $\vec j=(0,0,0,...,1)$, i.e., the term $(\Tr_F P^n)^{j_n}$ with $j_n=1$. Assuming that $n \neq 0, N,2N,3N,...$, then there is no $\bigO{g^0}$ term, and thus we have:
\begin{widetext}
\begin{eqnarray}
\nonumber
\langle {\cal P}_{\cal R}(\bm 0) {\cal P}_{\cal R}^\dagger(\bm r)\rangle &=& \frac{ n^2 g^4}{16 \pi^2}\chi_{\cal R}^{\mathbb I }\langle B_n(\bm 0) \cdot B^\dagger_n(\bm r) \rangle + \bigO{g^6} \\
	&=& \frac{ n^2 g^4}{16 \pi^2}\chi_{\cal R}^{\mathbb I }\sum_{j,l=1}^N  \Tr_F \left[ \Omega_0^n H_j \right] \Tr_F \left[ \Omega_0^{-n} H_l \right] \langle {b}_j(\bm 0)  {b}_l(\bm r) \rangle + \bigO{g^6}\,,
	\label{final expression of the correlator}
\end{eqnarray}
\end{widetext}
where $\chi_{\cal R}^{\mathbb I}\equiv\chi_{\cal R}\left(\vec j=(0,0,0,...,1)\right)$ and we have used the $\mathbb R^N$ basis in writing the double sum in (\ref{final expression of the correlator}). As we will see next, despite the fact that the pre-factor depends on the representation ${\cal R}$, the rest of this expression is a function of $\Tr_F \left[ \Omega_0^n H_j \right]$, which only depends on the N-ality of ${\cal R}$ since $ \Omega_0^n= \Omega_0^{\mod{n}{N}}$. 

The case $n = \mod{0}{N}$ (e.g. the adjoint) gives a term $\bigO{g^0}$, which leads to the behavior of the correlator $\langle {\cal P}_{\cal R}(\bm 0) {\cal P}_{\cal R}^\dagger(\bm r)\rangle\sim \mbox{constant}+{\cal O}(g^2)$. The ${\cal O}(g^0)$ term is interpreted as the breaking of the flux tube. This is the expected behavior of all zero N-ality representations since the probe charges of these representation can be completely screened by soft gluons. The breaking of adjoint strings are extremely difficult to see in lattice simulations, see e.g., \cite{Campbell:1985kp,Poulis:1995nn}. The continuity conjecture, on the other hand, provides a neat way to seeing the breaking.

\subsection{Combining everything: the string tension}

So far we have set the stage to finally obtain a closed-form expression of the Polyakov's loop correlator. First we recall that the Cartan generators $H_i$ are the components of the weights in the fundamental representation (defining representation), i.e. $H_i=\mbox{diag}\left((\nu_1)_i,(\nu_2)_i,...,(\nu_N)_i\right)$. Then, substituting (\ref{real space correlator}) into (\ref{final expression of the correlator}) we obtain
\begin{widetext}
\begin{eqnarray}
\nonumber
\langle {\cal P}_{\cal R}(\bm 0) {\cal P}_{\cal R}^\dagger(\bm r)\rangle&=&\frac{ n^2 g^4}{16 \pi^2N}\chi_{\cal R}^{\mathbb I}\sum_{j,l,k,m=1}^N\sum_{p=1}^N e^{\frac{2\pi i n}{N}\left(\bm \nu_k-\bm\nu_m\right)\cdot \rho}\left(\nu_k\right)_j\left(\nu_m\right)_l e^{-\frac{2\pi i p}{N}(j-l)} \langle \tilde b_p(\bm 0) \tilde b_{-p}(\bm r)\rangle\,.\\
\label{intermed step}	
\end{eqnarray}
Recalling that in the $\mathbb R^N$ basis we have $(\nu_a)_i=\delta_{ai}-\frac{1}{N}$, and that $\bm\rho=\sum_{b=1}^{N-1}\bm\omega_b$, where
\begin{eqnarray}
\bm\omega_b=\sum_{a=1}^b \bm e_a-\frac{b}{N}\sum_{a=1}^N \bm e_a\,,
\end{eqnarray}
we find $\bm \rho\cdot \bm \nu_b=-b+\frac{N+1}{2}$. Using this information in (\ref{intermed step}) we find three main terms that come from the multiplication $(\nu_k)_j(\nu_m)_l$:
\begin{enumerate}
\item The constant term $\frac{1}{N^2}$, which is the constant part of $(\nu_k)_j(\nu_m)_l$. This term is multiplied by the sum $\sum_{m=1}^Ne^{\frac{i2\pi nm}{N}}$, which is zero.
\item The term $\frac{\delta_{kj}}{N}$. Again, this term is multiplied by the sum $\sum_{m=1}^Ne^{\frac{i2\pi nm}{N}}$, which is zero.
\item Finally, we have the term $\delta_{kj}\delta_{ml}$, which is the only term contributing a non-zero value to $\langle {\cal P}_{\cal R}(\bm 0) {\cal P}_{\cal R}^\dagger(\bm r)\rangle$:
\begin{eqnarray}
\nonumber
&&\sum_{j,l,k,m=1}^N\sum_{p=1}^N e^{\frac{2\pi i n}{N}\left(\bm \nu_k-\bm\nu_m\right)\cdot \rho}\delta_{kj}\delta_{ml} e^{-\frac{2\pi i p}{N}(j-l)} \langle \tilde b_p(\bm 0) \tilde b_{-p}(\bm r)\rangle\\
\nonumber
&&=\sum_{p,k,l}e^{-\frac{2\pi i k(n+p)}{N}}e^{\frac{2\pi i l(n+p)}{N}}\langle \tilde b_p(\bm 0) \tilde b_{-p}(\bm r)\rangle=N\sum_{p,l}^N\delta_{n+p=0}e^{\frac{2\pi i l}{N}(n+p)}\langle \tilde b_p(\bm 0) \tilde b_{-p}(\bm r)\rangle\\
&&=N^2\langle \tilde b_p(\bm 0) \tilde b_{-p}(\bm r)\rangle_{p=-\mod{n}{N}}
\end{eqnarray}
\end{enumerate} 
Therefore, we finally obtain
\begin{eqnarray}
\langle {\cal P}_{\cal R}(\bm 0) {\cal P}_{\cal R}^\dagger(\bm r)\rangle= \frac{ N n^2 g^4}{16 \pi^2}\chi_{\cal R}^{\mathbb I} \langle \tilde b_k(0) \tilde b_{-k}(\bm r)\rangle_{k=\mod{n}{N}}\,.
\label{the main result}
\end{eqnarray}
\end{widetext}
Eq. (\ref{the main result}) is the main result of this work. It shows that apart from a nonuniversal and representation dependent pre-factor, the Polyakov-loop Correlator can only depend on the N-ality of representation. 

 We can use Eq. (\ref{the main result}) to obtain the string tension as follows. We are interested in length scales, $r>M_{\tilde b'_{k=\mod{n}{N}}}$, which is much bigger than the compactification length $L$. Therefore,  we  take the limit $r\rightarrow \infty$ in (\ref{p space correlator}):
\begin{eqnarray}
\nonumber
 &&\mbox{lim}_{r\rightarrow \infty}	\log \langle {\cal P}_{\cal R}(\bm 0) {\cal P}_{\cal R}^\dagger(\bm r)\rangle\\
&&=\mbox{constant}-{\cal M}_{\tilde b'_{k=\mod{n}{N}}}r\,,
\label{string tension and nality}
\end{eqnarray}
from which we read the string tension
\begin{eqnarray}
\sigma_k=L^{-1}{\cal M}_{\tilde b'_{k=\mod{n}{N}}}\,.
\end{eqnarray}
Thus, the string tension of the representation ${\cal R}$ will only depend on $\mod{n}{N}\neq 0$, which is the N-ality of the representation.

\section{W-bosons on the string worldsheet}
\label{W-bosons on the string worldsheet}

%
\begin{figure}[t]
\begin{center}
\includegraphics[width=85mm]{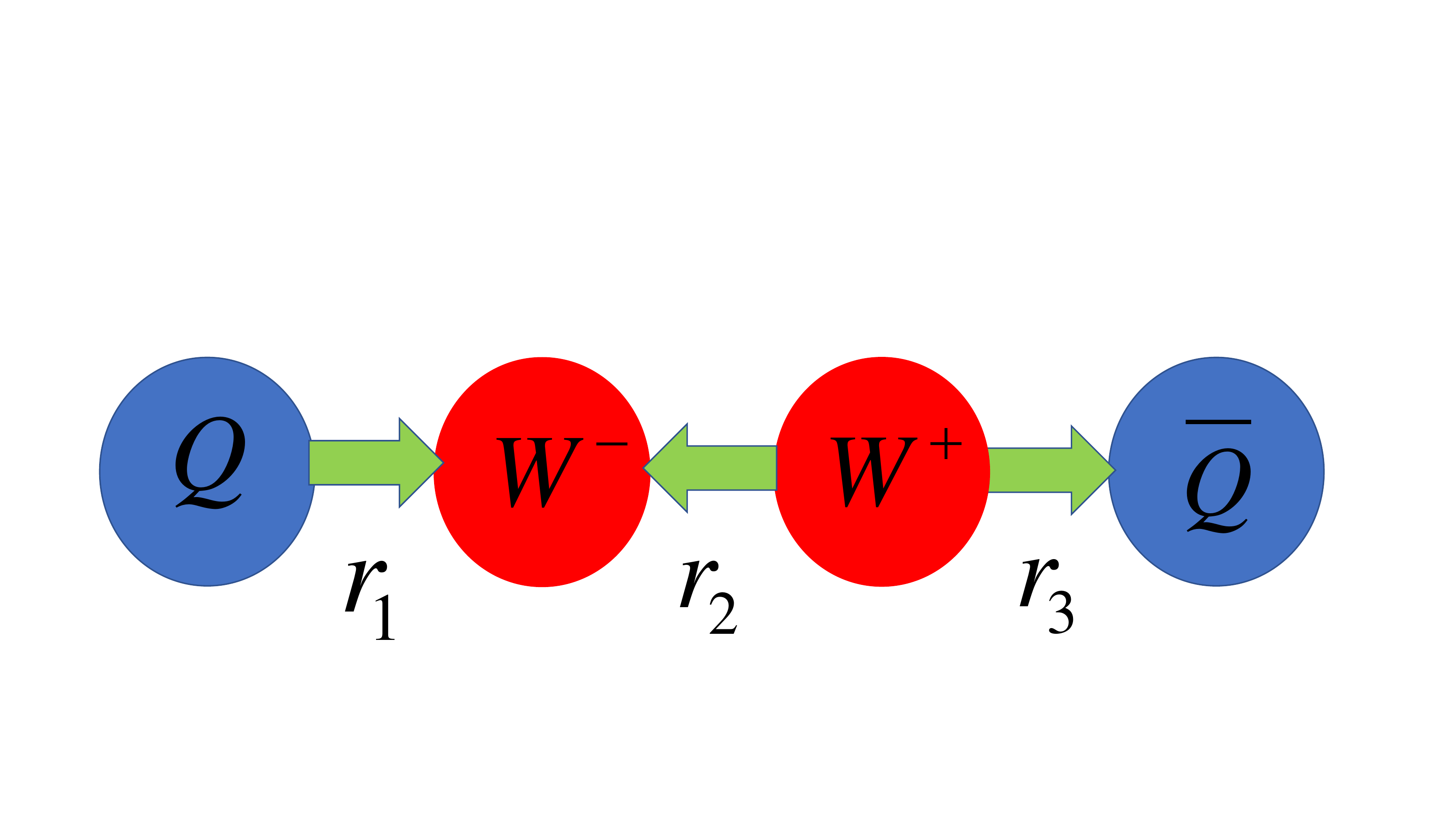}
\caption{W-bosons on the worldsheet of the ${\cal S}_{\mathbb R^3}$ string. In this specific example, the quarks, $Q,\bar Q$, are taken in the fundamental representation of $su(2)$. Since the W-bosons are in the adjoint representation, they carry twice the charge of the fundamental quark. The shown configuration is the only one that satisfies the conservation of the electric flux. $r_{1,2,3}$ label the W-bosons positions on the worlsheeet.}
\label{W bosons on worldsheets}
\end{center}
\end{figure}

 As a corollary of our main  result, Eq. (\ref{the main result}) one can also examine the effect of W-bosons on the string  between two probe charges in representation ${\cal R}$. We repeat our previous analysis in super Yang-Mills on a small circle by computing  correlators of Polyakov's loops wrapping the $\mathbb S^1$ circle with W-bsosns insertions. In the semi-classical limit the W-bosons are heavy and  we can neglect their kinetic energies.  They are charged under the moduli fields\footnote{See \cite{Anber:2013doa} for details.}, $\bm b$, (the charges live in the root system), and hence, they can exchange quanta of $\bm b$ with the probe charges. Therefore,  W-bosons can be thought of as adjoint Polyakov's loops  wrapping the circle and their effect on the string can be inferred by computing  higher Polyakov-loop correlators.   A typical correlator that is invariant under charge conjugation takes the form
\begin{eqnarray}
\nonumber
{\cal C}(\bm r,\bm r_1,\bm r_2)&=&\langle \Tr_{\cal R}\Omega(\bm 0)\Tr_{adj}\Omega_W(\bm r_1)\\
&&\times\Tr_{adj}\Omega_W^\dagger(\bm r_2)  \Tr_{\cal R}\Omega^\dagger(\bm r)\rangle\,,
\end{eqnarray}
and we assume that the N-ality of $\cal R$ is $k \neq 0$.  Since the W-bosons are in the adjoint representation, we  have
\begin{eqnarray}
\nonumber
\Tr_{adj}\Omega_W(\bm r_1)\cong -1+i\frac{g^2}{4\pi}\Tr_{adj}\left[e^{i \bm H\cdot \bm \Phi}\bm H\cdot \bm b\right]+{\cal O}(g^4)\,.\\
\label{trace of adjoint}
\end{eqnarray}
By assumption, the N-ality of  $\cal R$ is not zero, and hence, the expansion of $\Tr_{\cal R}\Omega$ starts at ${\cal O}(g^2)$. Then, the leading order contribution to  $C(\bm r,\bm r_1,\bm r_2)$ comes from  the first term in (\ref{trace of adjoint}) and ${\cal O}(g^2)$ term of $\Tr_{\cal R}\Omega$. Using (\ref{the main result}), we find that the correlator, to ${\cal O}(g^4)$, is given by
\begin{widetext}
\begin{eqnarray}
{\cal C}(\bm r,\bm r_1,\bm r_2)=\langle {\cal P}_{\cal R}(\bm 0) {\cal P}_{\cal R}^\dagger(\bm r)\rangle=\frac{ N n^2 g^4}{16 \pi^2}\chi_{\cal R}^{\mathbb I} \langle \tilde b_k(0) \tilde b_{-k}(\bm r)\rangle_{k=\mod{n}{N}}\,.
\end{eqnarray}
\end{widetext}
This shows that the N-ality of the string does not change by placing W-bosons on the string worldsheet. Also, the string tension is unaffected, to leading order in $g$, by the presence of W-bosons. The fact that the string tension does not get contribution from the W-bosons leads us to conclude that they are deconfined on the string worldsheet.

This result  was also reached in \cite{Anber:2015kea} by analyzing the ${\cal S}_{\mathbb R^3}$ strings on $\mathbb R^3$. Here, we provide a simple explanation of this interesting phenomenon. Let us consider two fundamental probe charges (quarks) of $su(2)$, $Q$ and $\bar Q$,  with opposite charges separated a distance $r$ and ending on the opposite sides of our ${\cal S}_{\mathbb R^3}$ string. The total energy of the system is $E=2m_Q+Tr$, where $m_Q$ is the quark  mass and $T$ is the ${\cal S}_{\mathbb R^3}$ string tension. The force between the quarks is $F=-dE/dr=-T$ and hence they experience linear confinement. Now, consider the same situation but with two W-bosons placed on the string worldsheet. Since the W-bosons belong to the adjoint representation, and hence they carry twice the charge of a fundamental quark, it is easy to convince oneself that the only configuration that respects the flux conservation is as shown in FIG. \ref{W bosons on worldsheets}.  The total energy of the system is $E=2m_Q+2m_W+T(r_1+r_2+r_3)$, where $m_W$ is the W-boson mass. Fixing the distance between the probe charges to be $r_1+r_2+r_3=r=\mbox{constant}$, we find that placing the W-bosons anywhere on the string worldsheet cannot change the energy of the system. Hence, the W-bosons  do not experience any force on the string worldsheet despite the fact that they interact logarithmically off the string\footnote{Super Yang-Mills on $\mathbb R^3\times \mathbb S^1$ is dimensionally reduced to $\mathbb R^3$. Electric charges in a  three-dimensional theory experience logarithmic interactions.}. 
 
Therefore, we learn from the above treatment of the W-bosons on ${\cal S}_{\mathbb S^1}$ and ${\cal S}_{\mathbb R^3}$ that they are deconfined on the worldsheets (experience no force) and they do not affect the string tension.  On the pure Yang-Mills side, the W-bosons are the soft gluons that cannot screen the non zero N-laity probe charges. This is a very intuitive phenomenon that is hardly proven in the strongly coupled regime. Nevertheless, we have shown that this phonomenon can be rigorously proven in the mass deformed super Yang-Mills on $\mathbb R^3\times \mathbb S^1$ and by continuity we conclude that the same phenomenon takes place in pure Yang-Mills theory.

\section{Discussion}
\label{Discussion}

In this work we have shown that the tension of the string wrapping the $\mathbb S^1$ circle depends, to leading order, on the N-ality of the representation. The next to leading order effect depends on the representation ${\cal R}$, and is expressed in terms of the group characters of the permutation group in representation ${\cal R}$. These findings exactly match holographic computations that were performed in the 't Hooft large-N limit \cite{Armoni:2006ri}.   It is extremely important to emphasize the role of center symmetry and the affine monopole in arriving to this result. The affine root is the way the theory remembers its four dimensional origin, and including the corresponding monopole is crucial to link super Yang-Mills to pure Yang-Mills via the conjectured continuity. 

 In terms of the strong coupling scale and the mass of the W-boson, the $k$-string tension is given by
\begin{eqnarray}
\nonumber
\sigma_k&=&\frac{\sqrt{81}}{\pi}\frac{N\Lambda^3}{m_W}\log^2\left(\frac{m_W}{\Lambda}\right)\\
&&\times\sin^2\left(\frac{\pi k}{N}\right)\sqrt{1-\frac{4\pi m m_W^2}{3\Lambda^3}\sin^{-2}\left(\frac{\pi k}{N}\right)}\,,
\label{string tension in terms of fundamental scales}
\end{eqnarray}
$k=1,2...,N-1$.  At small values of $m$ (this is the regime that is continuously connected to pure Yang-Mills theory),  the string tension $\sigma_n$ follows a square sine law:
\begin{eqnarray}
\frac{\sigma_k}{\sigma_1}=\frac{\sin^2\left(\frac{\pi k}{N}\right)}{\sin^2\left(\frac{\pi}{N}\right)}\,,
\label{square sine law}
\end{eqnarray}
where $\sigma_1$ is the fundamental string tension. This is in contradistinction with the Casimir law, $\sigma_k=\left(1-\frac{k-1}{N-1}\right)\sigma_1$, or sine law, $\sigma_k=\frac{\sin\left(\frac{\pi k}{N}\right)}{\sin\left(\frac{\pi}{N}\right)}\sigma_1$,  which have been advocated in the literature as two possible scalings of k-strings in Yang-Mills theories; see e.g. \cite{DelDebbio:2001kz,Lucini:2001nv,Lucini:2004my,Auzzi:2008ep}. The sine law in particular is consistent with 't-Hooft large-$N$ limit which requires the next to leading order correction of $\sigma_k$ to go as $1/N^2$ instead of $1/N$, as the Casimir law predicts. It is also consistent with various supersymmetric gauge theories and AdS/CFT computations; see e.g. \cite{Hanany:1997hr,Herzog:2001fq,Douglas:1995nw,Armoni:2003nz}.

Another question concerns the large-N limit of (\ref{string tension in terms of fundamental scales}), which has to be taken with care.  In the standard 't Hooft limit one takes $N \rightarrow \infty$ keeping $Ng^2$ fixed. In this limit the W-bosons of super Yang-Mills on $\mathbb R^3\times \mathbb S^1$ become very light,  $m_W\sim 1/(NL)$, which pushes the theory to strong coupling and invalidates the semi-classical treatment.  The proper limit in gauge theories on a circle is the abelian Large-N limit, which amounts to taking $N \rightarrow \infty$ keeping the W-boson mass fixed. In this limit we have $\sigma_k=k^2+{\cal O}\left(\frac{1}{N^2}\right)$, which is different from the expected 't Hooft large-N limit  $\frac{\sigma_k}{\sigma_1}=k+{\cal O}\left(\frac{1}{N^2}\right)$ in non-compact Yang-Mills theory. In the latter theory, the linear dependence of the string tension on the N-ality $k$  indicates that the string is made of $k$ independent components that do not interact with each other, which is not the case for the ${\cal S}_{\mathbb S^1}$ strings in the compactified theory\footnote{It was also shown in \cite{Cherman:2016jtu} that super Yang-Mills on $\mathbb R^3 \times \mathbb S^1$ in the abelian Large-N limit flows to a gapless theory in $\mathbb R^4$, which indicates  that the large-L and abelian large-N limits do not commute.}. 

The square sine law scaling in super Yang-Mills on $\mathbb R^3 \times \mathbb S^1$ as well as the unexpected large-N behavior is attributed to the fact that the string ${\cal S}_{\mathbb S^1}$ is much thicker than the compactification radius, and therefore, one should not expect that the string is composed of $N$ non-interacting components, as in the  $4$-D 't Hooft large-N case.  One expects, however, that the string tension departs from the square sine law and approaches the sine law in the $\Lambda^{-1}\ll NL$ limit. Assuming that the continuity between super and pure Yang-Mills holds, then this will happen in a way that preserves the N-ality dependence of the representation.

Finally, we compare our findings to other string models in the literature. In particular, we compare our ${\cal S}_{\mathbb S^1}$ strings to the ${\cal S}_{\mathbb R^3}$ strings that were studied in deformed Yang-Mills theory  \footnote{Deformed Yang-Mills is a theory with massive adjoint fermions or double trace deformation \cite{Unsal:2008ch}.} on $\mathbb R^3 \times \mathbb S^1$ in \cite{Poppitz:2017ivi} and also to strings in the softly broken Seiberg-Witten (SW) theory \cite{Seiberg:1994rs,Seiberg:1994aj}. We start with SW theory, where the strings are abelian in nature and of Abrikosov-Nielsen-Olesen type. The Weyl group in SW theory is broken, and therefore, one has $N-1$ different flux tubes corresponding to the $N-1$ fundamental weights $\bm \omega_a$, $a=1,2,..,N-1$, as was indicated in \cite{Douglas:1995nw}. The breaking of the Weyl group results in having different string tensions between quarks belonging to the same representation, depending on the specific weights of the quarks. For example, in $su(3)$ we have two non-degenerate strings $\bm \omega_1$ and $\bm \omega_2$, corresponding to the two fundamental weights. Hence, fundamental quarks (anti-quarks) with weights $\bm\nu_1,\bm\nu_2,\bm\nu_3$ ($\bar {\bm\nu}_1,\bar {\bm\nu}_2,\bar{\bm \nu}_3$) will have strings $\bm\mu_1,\bm\mu_2-\bm\mu_1,\bm\mu_2$, respectively. This is in contradistinction with ${\cal S}_{\mathbb R^3}$ strings in deformed Yang-Mills (dYM) theory, where we have unbroken Weyl group. This results in degenerate string tensions between all the fundamental quarks. For higher N-ality, the string tensions of a representation fall into distinct ${\mathbb Z}_N$ orbits, each of which has degenerate string tensions. In this regard, the ${\cal S}_{\mathbb R^3}$ strings of dYM are closer in nature to the QCD strings than the SW strings. The string tension in dYM, however, will in general depend on the representation, not only on its N-ality. For example, the two-index symmetric and two-index antisymmetric representations have different string tensions.  Unlike both types of strings (SW and dYM), we find that ${\cal S}_{\mathbb S^1}$ strings in super Yang-Mills, albeit the theory is still in the abelian regime,  depend only on the N-ality of the representation, making them identical to what one expects for QCD. 

This work lends extra support to the continuity picture between a class of deformed Yang-Mills theory on $\mathbb R^3 \times \mathbb S^1$ and real world QCD, including the conjectured continuity between super and pure Yang-Mills. Until now, there have been several tests to check the nature of this continuity, its regime of validity, and we were able to extract important  lessons about the four dimensional theory \cite{Aitken:2017ayq,Anber:2015wha,Anber:2017rch,Bhoonah:2014gpa,Anber:2017pak}.  It has been found that the deformed theories share a range of characteristics that point out to an underlying structure in the four dimensional Yang-Mills, which is not yet understood but is similar to the structure of the deformed theory.

\acknowledgements
We would like to thank Erich Poppitz and Mithat \"{U}nsal for  discussions. The work of M.A. is supported by the NSF grant PHY-1720135 and by Murdock Charitable Trust. The work of V.P. was supported in part by the Swiss National Science Foundation. 

\bibliographystyle{apsrev4-1}
\bibliography{k_strings_references}

\end{document}